\documentclass[aps,prl,twocolumn,superscriptaddress]{revtex4-1}

\usepackage{dcolumn}
\usepackage{bm}

\usepackage[T1]{fontenc}       
\usepackage{graphicx} 
\usepackage{epstopdf}
\usepackage{color}
\usepackage{amsmath}
\usepackage{amssymb}
\usepackage{float}
\usepackage[colorlinks]{hyperref}
\usepackage[absolute]{textpos}

\begin{document}

\author{Giang Thi Vu}
\email{gianthvu@uni-mainz.de}
\affiliation{Institut f\"{u}r Physik, Johannes Gutenberg Universit\"{a}t Mainz
Staudinger Weg 7, D-55099 Mainz, Germany}

\author{Anabella A. Abate}
\affiliation{Department of Physics, Universidad Nacional del Sur
- IFISUR CONICET, 800, Bahia Blanca, Argentina}

\author{Leopoldo R. G\'omez}
\affiliation{Department of Physics, Universidad Nacional del Sur
- IFISUR CONICET, 800, Bahia Blanca, Argentina}

\author{Aldo D. Pezzutti}
\affiliation{Department of Physics, Universidad Nacional del Sur
- IFISUR CONICET, 800, Bahia Blanca, Argentina}

\author{Richard A. Register}
\affiliation{Department of Chemical and Biological Engineering, Princeton University, Princeton, New Jersey, 08544, USA}

\author{Daniel A. Vega}
\email{dvega@uns.edu.ar}
\affiliation{Department of Physics, Universidad Nacional del Sur
- IFISUR CONICET, 800, Bahia Blanca, Argentina}

\author{Friederike Schmid}
\email{friederike.schmid@uni-mainz.de}
\affiliation{Institut f\"{u}r Physik, Johannes Gutenberg Universit\"{a}t Mainz
Staudinger Weg 7, D-55099 Mainz, Germany}

\title{Curvature as a guiding field for patterns in thin block copolymer films}

\begin{abstract}

Experimental data on thin films of cylinder-forming block copolymers (BC) --
free-standing BC membranes as well as supported BC films -- strongly suggest
that the local orientation of the BC patterns is coupled to the geometry in
which the patterns are embedded. We analyze this phenomenon using general
symmetry considerations and numerical self-consistent field studies of curved
BC films in cylindrical geometry. The stability of the films against
curvature-induced dewetting is also analyzed.  In good agreement with
experiments, we find that the BC cylinders tend to align along the direction of
curvature at high curvatures. At low curvatures, we identify a transition from
perpendicular to parallel alignment in supported films, which is absent in free-standing membranes. Hence both experiments and theory show that curvature 
can be used to manipulate and align BC patterns.

\end{abstract}

\maketitle

Because of their ability to self-assemble into well-defined periodic
nanostructures, block copolymers (BC) are attracting great interest
as potential template materials for cost-effective nanofabrication techniques
\cite{Harrison2000, Segalman2005, Bita2008, Ruiz2008, Singh2012, Marencic2010b,
Vega2013, Garcia2015, Abate2016}. With BC systems, one can produce
high-resolution patterns with tunable wavelength using traditional processing
techniques. This offers promising perspectives for applications in scalable
nanoscale devices.  However, one frequent problem with the self-assembly
approach is lack of long-range order due to pattern undulations and defects,
e.g., dislocations, disclinations, or grain boundaries \cite{Harrison2000,
Harrison2002, Vega2005,Nagpal2012,Hur2015}. Numerous methods to produce
patterns with well-defined orientational and positional order have
been proposed, such as shear alignment \cite{Angelescu2004,Kim2014b,Davis2015},
alignment in electric fields \cite{Amundson1993, Morkved1996,Mansky1998}, zone
annealing \cite{Berry2007, Yager2010}, or grapho- and chemo-epitaxy, where
surface interactions and confinement effects are exploited to order patterns
\cite{Kim2003,Gomez2009, Garcia2014, Garcia2015, Luo2015, Sundrani2004,
Yager2010, Marencic2010b, Hamley2009, Angelescu2004, Vega2013} or to
control defect positions \cite{Nelson2002}.

Here, we analyze another possible source of alignment, the {\em geometry} in
which the system is embedded.  Experiments and simulations on curved systems 
have indicated that the
pattern configurations are affected by both intrinsic and extrinsic geometry.
Even in Euclidean systems, a strong coupling between patterns and curvature
seems to drive the equilibrium configurations and the coarsening process
\cite{Gomez2009, Vega2013, Matsumoto2015, Pezzutti2015}.  

As a first step towards a more quantitative understanding of the nature of the
coupling between BC thin films or membranes and curvature, in the present
paper, we study curved monolayers of cylinder-forming BC systems by
complementary experiments, symmetry considerations, and self-consistent field
theory (SCFT) calculations \cite{Abate2016, Mueller2005}. We consider two types
of model systems: a) free-standing BC membranes and b) BC thin films deposited
onto a curved substrate. 

The geometric features of a 2D curved surface can be characterized in terms of
a shape operator ${\bf S}$, which has two Eigenvalues $k_{1,2}=1/R_{1,2}$
corresponding to the inverse maximal and minimal radii of curvature $R_i$
[see Supplemental Material (SM) for more details]. The
determinant and the trace of ${\bf S}$ define the Gaussian curvature $K=k_1
k_2$ and twice the mean curvature $2 H = k_1+k_2$, respectively \cite{Deserno}.
The experimental systems studied here have a non-Euclidean metric ($K \neq  0$,
free-standing membrane) or a Euclidean metric with zero Gaussian curvature ($K
= 0$, curved substrate).

In both systems we employ the same BC system, a cylinder-forming
polystyrene-block-poly(ethylene-alt-propylene) diblock (PS-PEP 4/13)
\cite{Marencic2010}.
The number-average block molecular weights for the BC are 4.3 kg/mol for PS and
13.2 kg/mol for PEP. In bulk, the PS blocks arrange in hexagonally packed
cylinders embedded in the PEP matrix. In thin films the PS cylinders adopt a
configuration parallel to the film surface.  The center-to-center spacing of
the cylinders is $d_{sm}=21$ nm.  Thin films of thickness $\sim 30$ nm are
prepared by spin-coating from a 1 wt. $\%$ solution in toluene, a good solvent
for both blocks.  Order is induced by annealing at a temperature $T$, above the
glass transition of the PS block ($T_g\sim 330$K) and below the order-disorder
transition temperature $T_{ODT}=417$K of the BC. Details of the
preparation of the experimental systems are given in SM. To obtain 
free-standing membranes, the films are first annealed on a flat
substrate, then further cooled down below the $T_g$ of
the PS block and finally lifted off and redeposited on a transmission electron
microscopy (TEM) grid.  During this process, the system retains the symmetry,
average inter-cylinder distance, and structure of defects established during
annealing. To obtain supported films, the BCs are directly spin-coated onto
curved substrates and the thermal annealing process is monitored by atomic
force microscopy (AFM) at selected time points.

\begin{figure}
\centering
\includegraphics[width=8.5cm]{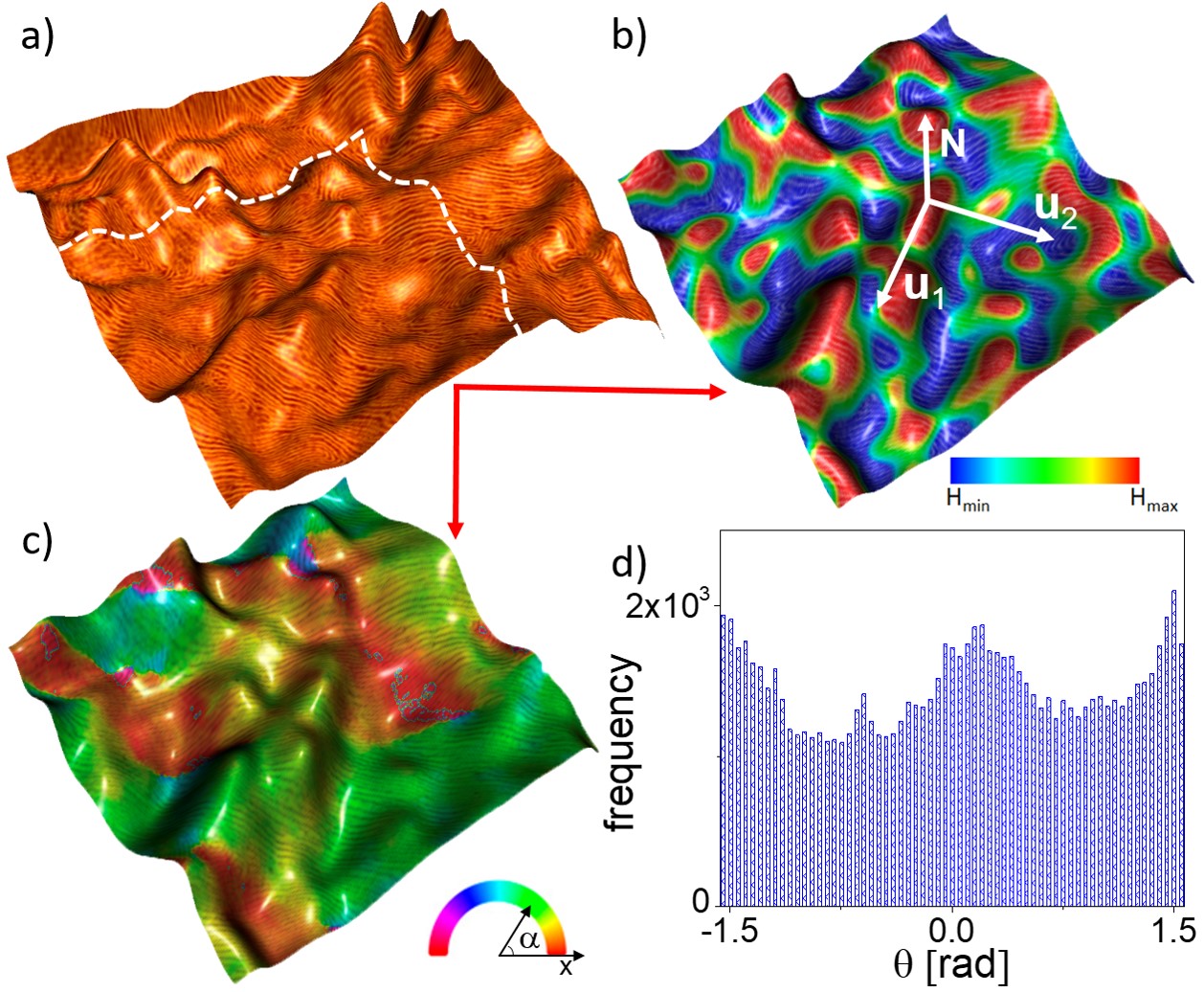} 
  \caption{\label{fig:fig1n} 
 a) Phase-height AFM image of a free-standing thin film (image
size: 2.6 $\mu$m $\times$ 2.6 $\mu$m). b) Local mean curvature for the membrane
shape.  Here the vectors $u_{1,2}$ indicate the directions of the principal
curvatures and $\textbf{N}$ is the normal vector to the membrane surface (image
size: 1.8 $\mu$m $\times$ 1.8 $\mu$m, region indicated by dashed lines in panel a;
$H_{max}=-H_{min}=3.84 \times 10^{-3}$ nm${}^{-1}$). c) Local orientation of
the director field $\alpha$ of the pattern with regard to the x-axis.  d)
Histogram showing the local distribution of angles $\theta=\alpha-\beta$
between $\alpha$ and the local orientation of the membrane wrinkles $\beta$
[see also Fig.\ \ref{fig:figSB} in SM].}
\end{figure}

Fig.\ \ref{fig:fig1n} shows an AFM image of a free-standing film where height
and cylinder locations are measured simultaneously.  The light and dark regions
correspond to PS-rich and PEP-rich regions, respectively. After releasing the
membrane from the confining substrate, it develops a non-Euclidean shape to
relieve the elastic energy of topological defects that have survived the
thermal annealing. The shape results from a competition between the strain
field of the defects, the bending energy associated with the curvature of the
membrane, and the membrane tension \cite{Matsumoto2015}. Fig.\
\ref{fig:fig1n} (d) shows the correlation between the orientation of the
underlying pattern and the local orientation of membrane wrinkles. Although
the different defects impose competing out-of-plane deformations, one clearly
notices that wrinkles have a tendency to be oriented either perpendicular
 ($\theta = 0$) or parallel ($\theta = \pm \pi/2$) to the underlying cylinders, 
 suggesting that the bending energy is
anisotropic and coupled to the liquid crystalline order of the BC
[see also Figs.\ \ref{fig:figS1}, \ref{fig:figS2} in SM].

\begin{figure}[t]
\centering
\includegraphics[width=8.cm]{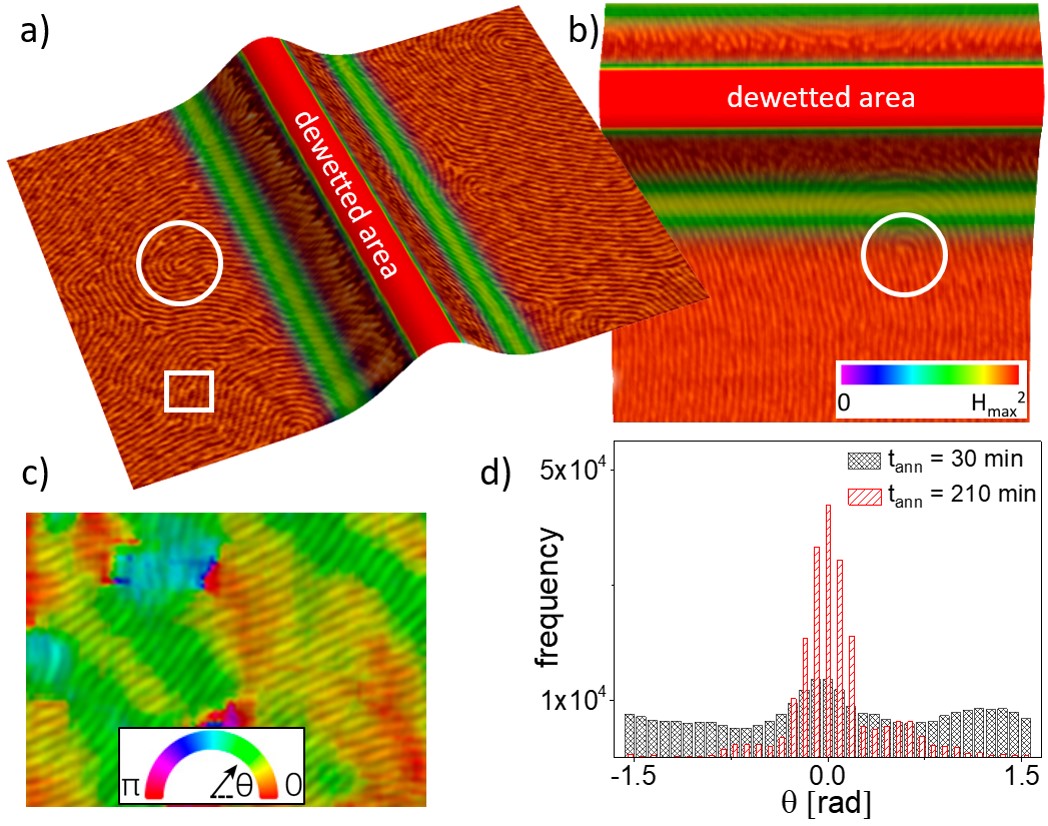}
  \caption{\label{fig:fig2n} 
Top panels: 3D AFM phase-height images of the BC thin film on a curved
substrate after annealing at T=373K. Panels a) and b) show the pattern
configuration after 90 min (image size: 2.0 $\mu$m $\times$ 1.5 $\mu$m) and 3.5
h of thermal annealing (image size: 1.0 $\mu$m $\times$ 1.25 $\mu$m),
respectively.  Height scale:  80 nm from crest to valley, $H_{max}^2 = 6.25$ $\mu$m${}^{-2}$). 
The presence of a dislocation and $+1/2$ disclinations has
been emphasized with a rectangle and circles, respectively. Bottom panels: c)
Local orientation of the smectic pattern (color map indicated at the bottom).
d) Histograms showing the distribution of angles $\theta$ between
the local cylinder orientation and the direction of curvature at two different
annealing times.
}
\end{figure}

A similar observation is made for the thin films on curved substrates. 
Fig.\ \ref{fig:fig2n} shows AFM phase and height-phase images of the BC thin
film deposited onto a curved substrate. Right after the spin coating, the
pattern is characterized by a very small orientational correlation length
($\xi_2 \sim 20 $ nm) and a high density of defects. During annealing at T=373
K, the system orders via annihilation of dislocations and disclinations.
Already at an early stage of annealing, the thin film becomes unstable and
dewets at the regions with the highest curvature (Fig.\ \ref{fig:fig2n}). Upon
further annealing, the order in the system increases and the pattern develops a
clear preferential orientation with regard to the substrate. Fig.\
\ref{fig:fig2n} shows that the PS cylinders tend to align perpendicular to the
crest of the substrate. Thus the topography of the substrate seems to act as an
external field that breaks the azimuthal symmetry [see also Fig.\
\ref{fig:figS3} in SM]. Note that the equilibrium configuration
obtained here is opposite to that predicted in previous theories for curved
columnar phases \cite{Santangelo2007, Kamien2009}, where it was assumed
that bending along the cylinder direction is energetically more costly
than bending in the perpendicular direction \cite{Kamien2009}.

The phenomena described above can be analyzed using general
symmetry considerations. The curvature free energy per area of {\em isotropic}
fluid-like membranes can be expanded in the invariants of the shape operator
${\bf S}$ as $F_{HC}=\frac{\kappa_b}{2} (2H-c_0)^2 + \kappa_g K$, where
$\kappa_b$ and $\kappa_g$ are the bending and Gaussian rigidity, respectively,
and $c_0$ is the spontaneous curvature \cite{Helfrich1973,Canham1970}. Here we
consider anisotropic {\em nematic} membranes with in-plane order characterized
by a director ${\bf n}$ (the orientation of the cylinders), thus additional
terms become possible.  Including all terms up to second order in
${\bf S}$ that are compatible with the in-plane nematic symmetry, i.e., $({\bf
n} \cdot {\bf S} \cdot {\bf n})$, $({\bf n} \cdot {\bf S} \cdot {\bf n})^2$
\cite{powers1995,biscari2006}, and $({\bf S} \cdot {\bf n})^2$
\cite{oda1999,chen1999}, 
we can derive the following expression for the anisotropic part of the
curvature free energy per area [see SM]:
\begin{equation} \label{eq:ani} 
  F_{\mbox{\tiny ani}} 
    = -\frac{\kappa'}{2} (k_1 - k_2) (2 H -c_0') \cos(2 \theta) 
      - \frac{\kappa''}{2} (H^2 - K) \cos(4 \theta) .
\end{equation} 
Here $\theta \in [0:\pi/2]$ denotes the angle between the director and the
direction of largest curvature $k_1$ ($|k_1| > |k_2|$), and $\kappa',
\kappa''$, $c_0'$ are anisotropic elastic parameters. In symmetric membranes,
$c_0'$ vanishes ($c_0'=0$). We emphasize that Eq.\
(\ref{eq:ani}) gives the generic form of the free energy of curved nematic
films up to second order in the curvatures, which should be generally valid
regardless of molecular details.  For $\kappa''>0$, the second term describes a
quadrupolar coupling between the curvature tensor and the director that favors
{\em two} directions of preferential alignment of the director ${\bf n}$ along
the two principal directions of curvature.  Such a competition between two
stable/metastable aligned states was also predicted in other continuum models
for nematic shells \cite{Napoli2012,Napoli2012b}. The first term selects
between the two directions. 

The results of the symmetry analysis are compatible with the
experiments: As discussed above, in the BC membranes, wrinkles form
preferentially parallel or perpendicular to the director [Fig.\ \ref{fig:fig1n}
(d)]. Similarly, in the thin films, the distribution of local cylinder
orientations $\theta$ is bimodal at early annealing time (30 min),  with two
characteristic peaks separated by $\sim \pi/2$ [Fig.\ \ref{fig:fig2n}(d)].
During the first stage of coarsening, the parallel and perpendicular
configurations compete.  After long annealing times, $C_{\perp}$ dominates, and
the histogram becomes sharply peaked at the orientation $\theta=0$, 
suggesting $\kappa'>0$ in Eq.\ (\ref{eq:ani}). We note, however,
that supported films are asymmetric and hence the spontaneous curvature
parameter $c_0'$ will very likely not vanish, in which case Eq.\ (\ref{eq:ani})
predicts that the preferred orientation switches from $\theta=0$ to
$\theta=\pi/2$ in a region of very small curvatures $k_1 \in [0:c_0']$. We will
discuss this further below.

\begin{figure} [t]
\centering
 \includegraphics[width=8.2cm]{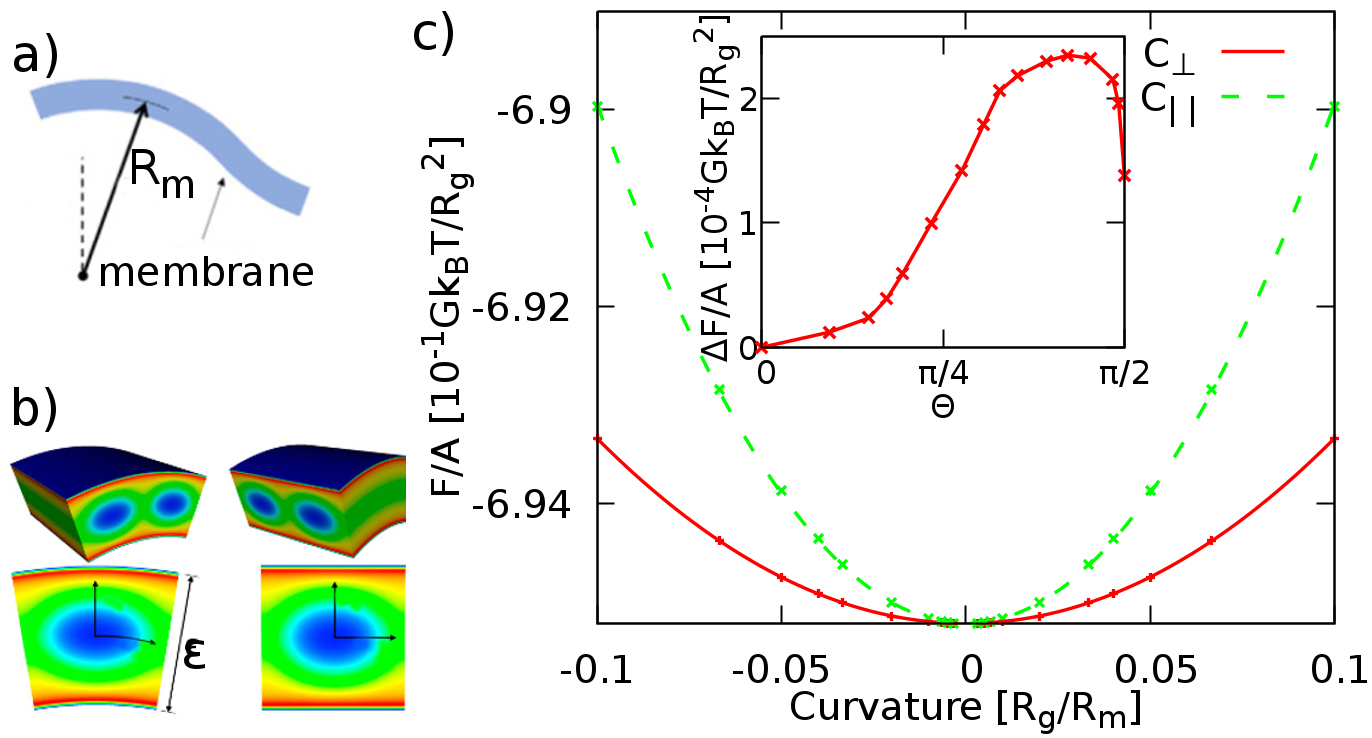}
  \caption{\label{fig:fig3n} (a) Schematic representation of curvature radius
$R_m$ in free-standing membranes. (b) Density profiles from SCFT for the
parallel (left) and perpendicular (right) configurations $C_\parallel$ and
$C_\perp$ at $R_m=9Rg$. (c) Free energy per area as a function of inverse
curvature radius $R_g/R_m$ for the $C_{||}$ and $C_{\perp}$ configurations.
Inset shows the free energy shift per area as a function of angle $\theta$ 
between the cylinders and the direction of curvature relative to the
$C_{\perp}$ configuration ($\theta = 0$) at $R_m = 50 R_g$.  }
\end{figure}

\begin{figure} [t]
\centering
 \includegraphics[width=7.8cm]{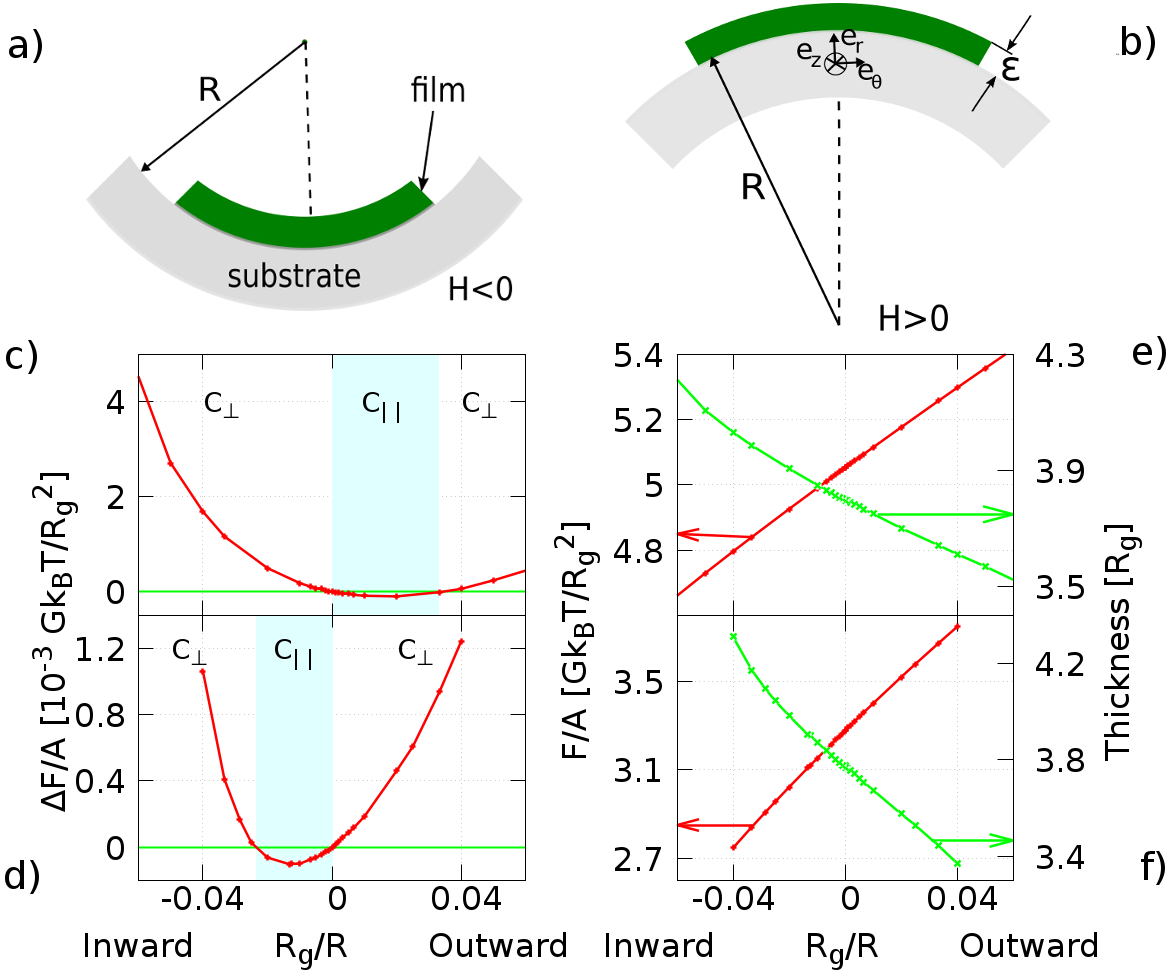}
 \caption{\label{fig:fig4n} (a,b) Schematic representation of supported
 thin films (green) on curved substrates (gray). 
 (c,d) Free energy difference per area $\Delta F = (F_{||} - F_{\perp})/A$ 
of supported thin films with parallel ($C_{||}$) vs. perpendicular 
($C_{\perp}$) orientation, versus curvature, for two sets of surface
interaction energies I (top) and II (bottom) as described in the main text.
Blue shaded regions highlight curvature regimes where the parallel 
orientation is more favorable ($\Delta F < 0$).
 (e,f) Corresponding curves for the free energy per area (red) 
 and thickness (green) in the perpendicular configuration $C_\perp$.}
\end{figure}

In order to obtain a more quantitative theoretical
description, we use SCFT \cite{Matsen2002,Mueller2005} to study the two systems
considered in the experiment, the free-standing membrane and the curved
supported thin film. We consider a melt of asymmetric $AB$ diblock copolymer
molecules with degree of polymerizaton $N$ and statistical
segment length $b$ at temperature $T$ confined to a curved film of thickness
$\epsilon$ by two coaxial cylindrical surfaces (see schematics in Fig.\
\ref{fig:fig3n}(a), \ref{fig:fig4n}(a), and \ref{fig:fig4n}(b). Periodic or tilted periodic boundary
conditions [see SM] are applied in the two in-plane directions. 

In the following, lengths and energies are given in units of $R_g^2 =
\frac{1}{6} N b^2$ and $G k_B T$, respectively, where $k_B$ is the Boltzmann
constant and $G =\rho_c \: R_g^3$ is the rescaled dimensionless copolymer
density in the bulk. The incompatibility between the blocks is specified by the
product $\chi N$, where $\chi$ is the Flory-Huggins parameter. Here we use
$\chi N=20$ and $f=0.7$ to match the experimental values ($f$ is the volume
fraction of the $A$-block). Our calculations are done in the grand canonical
ensemble with the chemical potential $\mu=(2.55 + \ln G) k_BT$
\cite{Abate2016,fn1} and inverse isothermal compressibility $\kappa N=25$
\cite{Helfand1975, Pike2009, Detcheverry2010}. Monomers $\alpha = A,B$ close to a surface
experience a surface field, which we characterize in terms of the surface
energy per area $\gamma_\alpha$ of a fluid of $\alpha$-monomers in contact with
the same surface [see SM], given in units $\hat{\gamma} = G k_B T/R_g^2$. To
account for the experimental fact that cylinders align parallel to the film,
the interaction parameters are chosen such that majority A-blocks
preferentially adsorb to the surface, i.e., $\gamma_A < \gamma_B$. In planar
films, the copolymers then self-assemble into aligned cylinders 
with a spacing $\lambda= 3.6 R_g$. Matching this with the value $d_{sm} = 21$
nm observed experimentally, we can identify $R_g \approx 5.8$ nm 
\cite{fn3} and hence $G = 5.77$ for our experimental systems (assuming an
average copolymer density of 0.861 g/cm${}^3$ at 363K).

We first consider free-standing membranes, which we model as a symmetric film
with surface interaction energies $\gamma_A N = - 24 \hat{\gamma}$ and
$\gamma_B N = - 23 \hat{\gamma}$. We calculate the free energy per area as a
function of the curvature radius $1/R_m$ of the mid-surface of the film (see
Fig.\ \ref{fig:fig3n}(a) for the two cases where cylinders are aligned parallel
or perpendicular to the curvature ($C_\parallel$, $C_\perp$, see Fig.
\ref{fig:fig3n}(b). In each case, the film thickness $\epsilon$ and the
wavelength of the characteristic pattern are optimized to obtain the lowest
free energy state. Fig. \ref{fig:fig3n}(b) shows the resulting density profiles
for the parallel and perpendicular configurations in a system with a relatively
large curvature ($R_m=9Rg$). The differences are small, indicating that
curvature affects neither the position of the cylinder with regard to the plane
of symmetry, nor the segregation strength. The optimum inter-cylinder spacing
is $\lambda \sim 3.6 R_g$, which is slightly smaller than the bulk value,
$\lambda_{bulk} \sim 3.7 R_g$. The ratio, $\lambda /\lambda _{bulk} \sim 0.97$,
is in good agreement with SCFT calculations and experiments on flat substrates,
where it was found that in thin films the unit cell is stretched perpendicular
to the plane of the film resulting in lateral distances smaller than those in
bulk \cite{Knoll2007,Abate2016}. The optimal thickness is $\epsilon \sim
3.5R_g$ for both the parallel and perpendicular configurations [see  Fig.\
\ref{fig:figS7} in SM].  None of these features appears
to be severely affected by the curvature within the range of curvatures
explored here.

The behavior of the free energy per area for the two configurations is shown in
Fig.\ \ref{fig:fig3n}(c). The perpendicular orientation is clearly favored.
Furthermore, in agreement with Eq.\ (\ref{eq:ani}), both
$C_{\perp}$ and $C_{||}$ represent local free energy minima with respect to
variations of the angle $\theta$ between cylinders and the direction of
curvature, (see inset of Fig.\ \ref{fig:fig3n}c).  Since the free energy grows
almost quadratically with the curvature, ${F}/{A}={\kappa}/{2R_m^2}$, we can
calculate bending stiffness parameters for the $C_{||}$  and $C_{\perp}$
configurations.  By fitting the free energy per area up to a quadratic order of
the mean curvature, we obtain $\kappa_{||}=(1.056 \pm 0.002)G k_BT$ and
$\kappa_{\perp}=(0.376\pm 0.002)G k_BT$ for the parallel and perpendicular
configurations, respectively.  Comparing this with Eq.\ (\ref{eq:ani}) and
using $k_1 = \pm 1/R_m$, $k_2 = 0$, we can deduce $\kappa'=(\kappa_{||} -
\kappa_\perp) = 0.68 G k_B T \approx 4 k_B T$, which corresponds to $\kappa'
\approx 1.6 \times \: 10^{-13}$ erg at room temperature. In contrast, the total
bending energy of the membranes has been estimated to be of order $\kappa_b
\sim 10^{-9}$ erg, which is much higher due to the large contribution of the
glassy PS block \cite{Matsumoto2015}. Hence, the influence of $\kappa'$ on the
membrane shapes is presumably negligible.

The situation is different when looking at copolymer ordering in supported
films, where the curvatures are kept fixed and energy differences of order $k_B
T$ do significantly influence the selection of the pattern orientations. 

Thin films differ from membranes in two respects. First, the
reference surface which is kept fixed during the free energy minimization is
the interface between the substrate and the film (not the mid-plane of the film
as in membranes), and second, the interaction energies may be different at the
substrate and air interfaces.  Fig.\ \ref{fig:fig4n}(c),\ref{fig:fig4n}(d) shows results for
$\gamma_B^{a,s}= -6 \hat{\gamma}$ and two representative
parameter sets for $\gamma_A^{a,s}$: (I) $\gamma_A^s N = -10 \hat{\gamma},
\gamma_A^a N = -20 \hat{\gamma}$, and (II) $\gamma_A^s N = -24 \hat{\gamma},
\gamma_A^a N = -10 \hat{\gamma}$, where superscripts $s$ and $a$ denote
``substrate'' and ``air'', respectively.  In both cases, the perpendicular
configuration is more favorable at large curvatures. At small curvatures,
however, there exists a small regime where parallel configurations have lower
free energy. This is in agreement with our symmetry
considerations (see above, Eq.\ (\ref{eq:ani})) and also with the experimental
observations.  Indeed, Fig.\ \ref{fig:fig2n}(b) and Fig.\ \ref{fig:figS1} in SM
suggest that the locally preferred orientation switches from perpendicular to
parallel in a region around the inflection point of the surface profile (green
shaded areas in Fig.\ \ref{fig:fig2n}), and this induces defects in that
region. Hence curvature can be used not only to orient patterns, but also to
generate defects at specific regions in space.

Fig. \ref{fig:fig4n}  also shows the behavior  of the free energy and the
minimum-energy thickness as a function of mean curvature for the $C_{\perp}$
configuration.  For $H\ge 0$, the free energy increases as the curvature
increases, indicating that the thin film is likely to become unstable and dewet
from the substrate, also in good agreement with the experimental data shown in
Fig. \ref{fig:fig2n}.  Conversely, for $H<0$ the film remains stable, since the
free energy decreases as the curvature increases.  In the experiments (Fig.
\ref{fig:fig2n}), it can be observed that the thin film dewets at the region
with the highest curvature, where $H_{max}R_g \sim 0.015$. 
These results are in good agreement with recent experiments on curved
substrates by Park and Tsarkova \cite{Park2017}, who also found dewetting for
$H>0$ and thin film thickening for $H<0$ in agreement with Fig.\
\ref{fig:fig4n}(e),\ref{fig:fig4n}(f) (green curves).

In conclusion, we have shown through experiments, symmetry considerations, and
SCFT calculations that curvature can be employed as a guiding field to produce
well-ordered patterns. The SCFT calculations provide a rough estimate of the
equilibrium configuration for curved systems and predict dewetting in regions
with high local positive curvature $H>0$. From a technological perspective, our
results indicate that through appropriate control over the surface
interactions, it should be possible to prevent dewetting while keeping a
geometric field with sufficient strength to guide order.

\section{Acknowledgements}

We gratefully acknowledge the financial support from the Deutsche
Forschungsgemeinschaft (Grant Schm 985/19 and SFB TRR 146), the National
Science Foundation MRSEC Program through the Princeton Center for Complex
Materials (DMR-1420541), the Universidad Nacional del Sur, and the National
Research Council of Argentina (CONICET). The SCFT calculations were done on the
high performance computing center MOGON in Mainz.  PS-PEP 4/13 was synthesized by Dr. Douglas Adamson.

%

\clearpage
\section{Supplementary Information on: \\
Curvature as a guiding field for patterns in thin block copolymer films}

This supplementary material provides additional information about the 
experimental systems and details on the theory and the SCFT calculations.

\setcounter{figure}{0} 
\renewcommand{\thefigure}{S\arabic{figure}} 
\setcounter{equation}{0} 
\renewcommand{\theequation}{S\arabic{equation}} 

\subsection{Experimental Systems}

\subsubsection{System Preparation}

The polystyrene-block-poly(ethylene-alt-propylene) diblock
(PS-PEP 4/13) copolymer was synthesized through sequential living anionic
polymerization of styrene and isoprene followed by selective saturation of the
isoprene block (see ref.  \cite{Marencic2010} for details). 
Thin films of thickness $\sim 30$ nm are prepared by spin-coating
from a 1 wt. $\%$ solution in toluene.

To obtain a free-standing membrane, we first spin coat a monolayer of BC
cylinders onto a 50 nm thick flat layer of sucrose deposited onto a silicon
wafer, then thermally anneal it at $T$=363K until a prescribed orientational
correlation length of $\xi_2 \sim 200$ nm is obtained \cite{Matsumoto2015}, and
finally cool it to room temperature. The sucrose layer is then used as a
sacrificial layer to float the thin film off the substrate and onto the surface
of water, and subsequently redeposit it as a free-standing membrane on a
transmission electron microscopy (TEM) grid (grid spacing $25 \, \mu$m $> \xi_2 >
d_{sm}$). As the film is lifted off at a temperature well below the glass
transition temperature of the PS block, the system retains the symmetry,
average inter-cylinder distance, and structure of defects established during
annealing. The pattern order is mainly disrupted by $\pm \frac{1}{2}$
disclination multipoles.  

To prepare the substrate, we employ a solvent-annealing technique on a
photoresist array of trench patterns deposited onto a silicon nitride wafer.
Details about the method of sustrate preparation can be found elsewhere
\cite{Vega2013}. It yields Gaussian-like smooth substrates with a pitch of 2.2
$\mu$m and crests with maximum height of 80 nm. The largest mean curvature of
the substrate is found at the crests, where $H=H_{max}\sim 2.5 \,
\mu$m${}^{-1}$. 

The thin films are imaged using a Veeco Dimension 3000 atomic force microscope
(AFM) in tapping mode. The spring constant of the tip (uncoated Si) is $\sim$
40 N/m and its resonant frequency is 300 kHz. 

\subsubsection{Membrane topography}

In order to determine the local curvatures, the metric tensor for the membrane
was obtained at room temperature from the AFM height profiles.
Through AFM we parametrize the surface in the Monge gauge. In this
representation, the coordinates of each point $\textbf{r}$ are expressed as
$\textbf{r}=(x,y,h(x,y))$, where $x$ and $y$ are planar coordinates and
$h(x,y)$ the out-of-plane displacement. The metric tensor can then be
calculated as $g_{ij} = \delta_{ij} + h_i h_j$ and the shape tensor as

\begin{eqnarray}
{\bf S} & = & \frac{1}{(1+h_x^2 + h_y^2)^{3/2}} 
\\ \nonumber &&
\left( \begin{array}{cc} 
(1+h_y^2) h_{xx} - h_x h_y h_{xy} & (1+h_x^2) h_{xy} - h_{xx} h_x h_y \\
(1+h_y^2) h_{xy} - h_x h_y h_{yy} & (1+h_x^2) h_{xy} - h_{xy} h_x h_y 
\end{array} \right).
\end{eqnarray}
The Eigenvalues of ${\bf S}$ give the principal curvatures.  Standard
methods are employed to determine the metric tensor and the principal
curvatures at each point on the membrane surface \cite{Matsumoto2015}.

\begin{figure}[h]
\includegraphics[width=8.5cm]{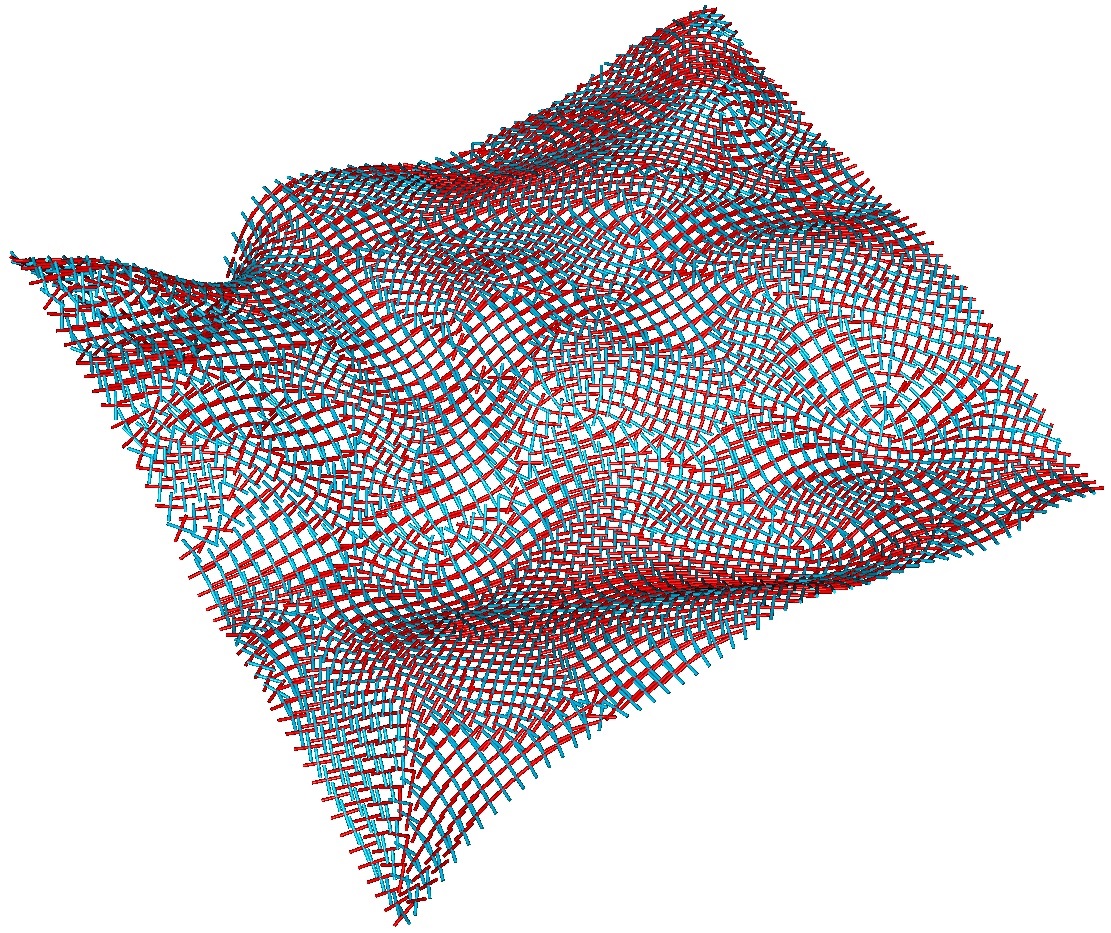}
  \caption{\label{fig:figSA} {Unit vectors indicating the directions of the principal maximum (red lines) and minimum (light blue lines) curvatures.}}
\end{figure}

\begin{figure*}
\includegraphics[width=15cm]{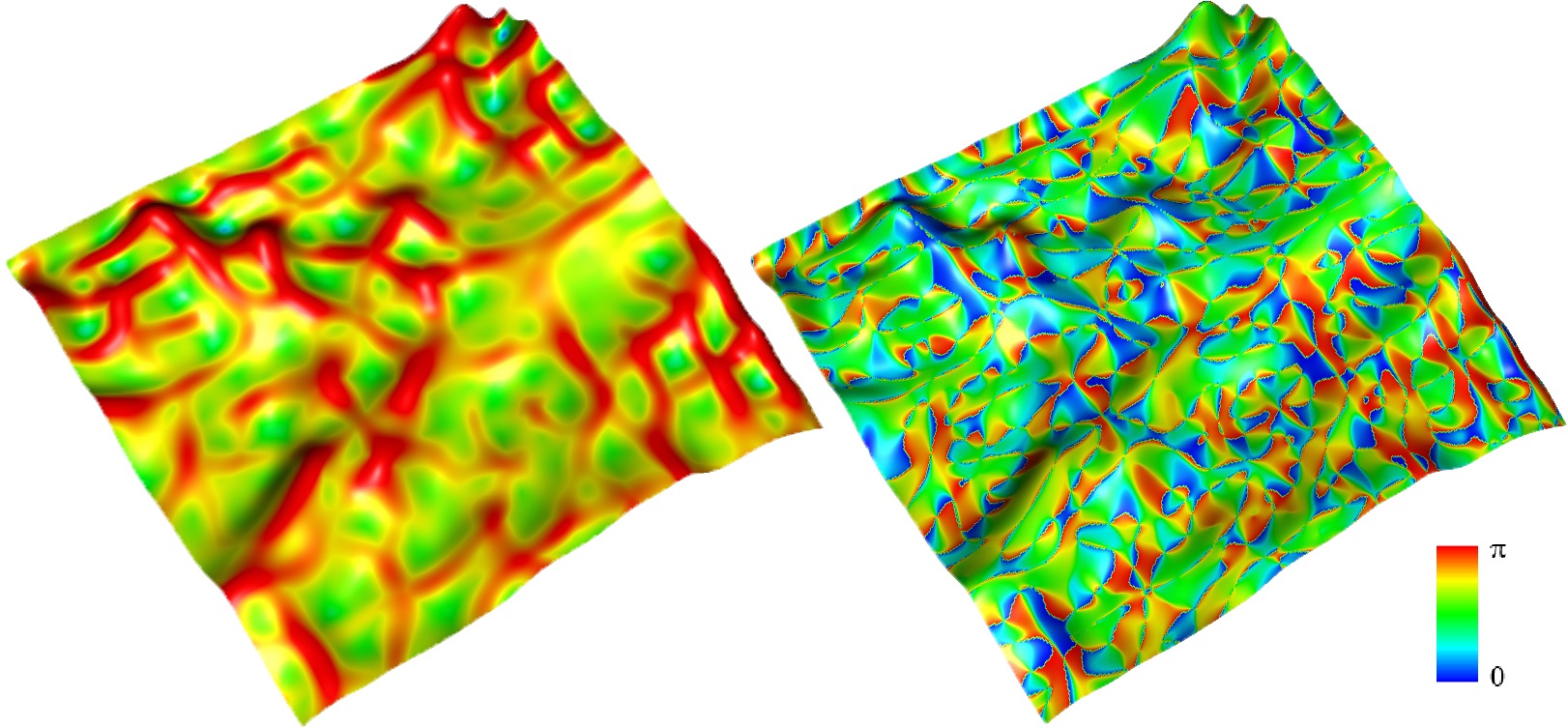}
  \caption{\label{fig:figSB} Local values of maximum principal curvature allow clear identification of the location and local orientation $\beta$ of the membrane wrinkles. 
  Left panel: maximum principal curvature. Right panel: orientation $\beta$ of the wrinkles with regard to the x-axis.}
\end{figure*}

Figure\ \ref{fig:figSA} shows the direction of the principal
minimum and maximum curvatures. Once these directions and the principal radii
of curvature $R_{1,2}$ have been determined, the geometric properties of
the membrane shape can be obtained.  To determine the correlation between the
pattern orientation $\alpha$ and membrane distortions,  we determine the local
orientation of the membrane wrinkles $\beta$ relative to the x-axis from
the distribution of the maximum main curvature $k_1$. The wrinkle orientation
$\beta$ was obtained by measuring the local gradient of $k_1$ (see Fig.
\ref{fig:figSB}), i.e.  $\tan{(\beta)}=\nabla_y K_1 / \nabla_x K_1$.

\begin{figure}[h]
\includegraphics[width=8.5cm]{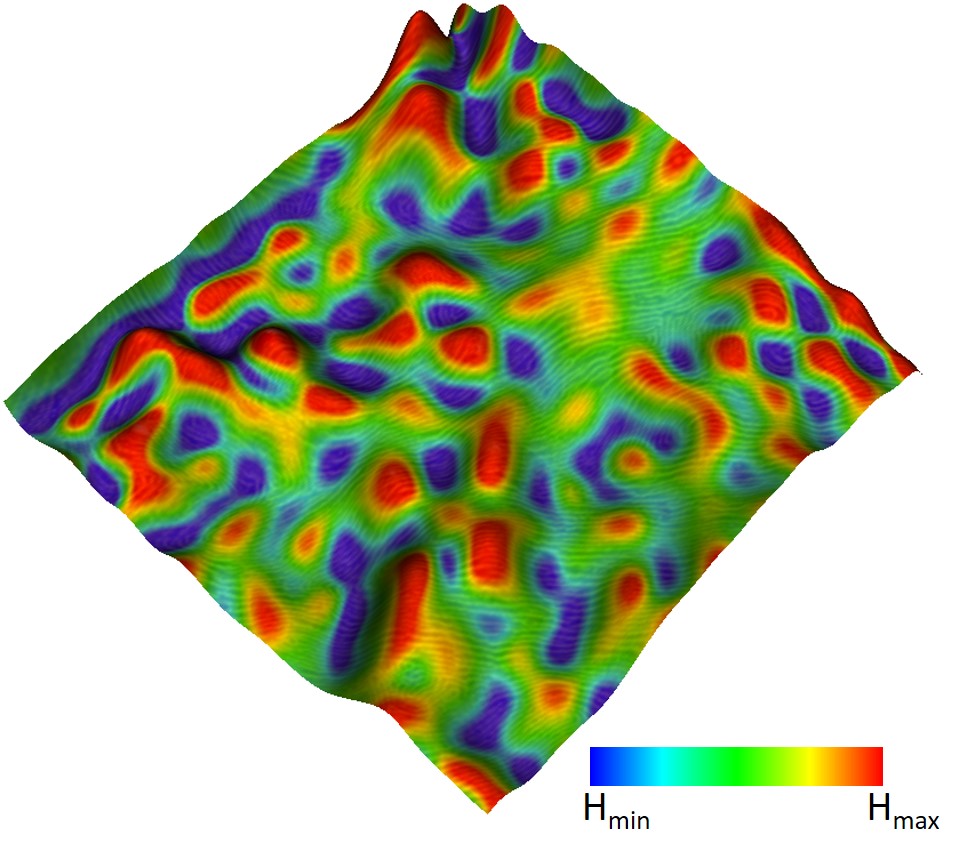}
  \caption{\label{fig:figS1} Height-phase AFM image of a freestanding thin film
overlapped with mean curvature. ($H_{max}=-H_{min}=3.84 \times 10^{-3}$
nm${}^{-1}$; image size: 2.6 $\mu$m $\times$ 2.6 $\mu$m).}
\end{figure}

\begin{figure}[h]
\includegraphics[width=8.5cm]{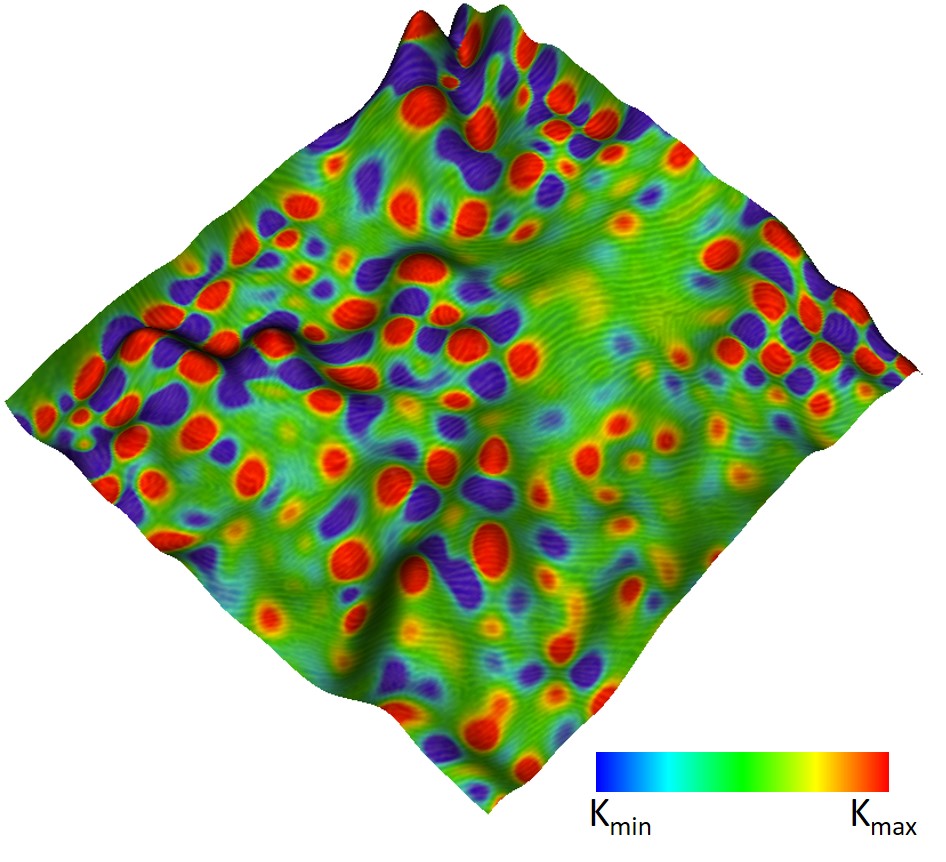}
  \caption{\label{fig:figS2} Height-phase AFM image of a freestanding thin film
overlapped with Gaussian curvature. ($K_{max}=-K_{min}=1.2 \times 10^{-5}$
nm${}^{-2}$; image size: 2.6 $\mu$m $\times$ 2.6 $\mu$m).}
\end{figure}

Figs.\ \ref{fig:figS1} and \ref{fig:figS2} show the mean and Gaussian
curvatures  for a free-standing cylinder-forming BC thin film membrane. 
Here a semitransparent mask with the maps for $H$ and $K$ was applied to also show the block copolymer texture (70$\%$ transparency).

Note the wrinkled topography of the membrane and the coupling with the smectic-like
texture of the BC system.

\subsubsection{Thin films on curved substrates}

\begin{figure}[h]
\centering
\includegraphics[width=8cm]{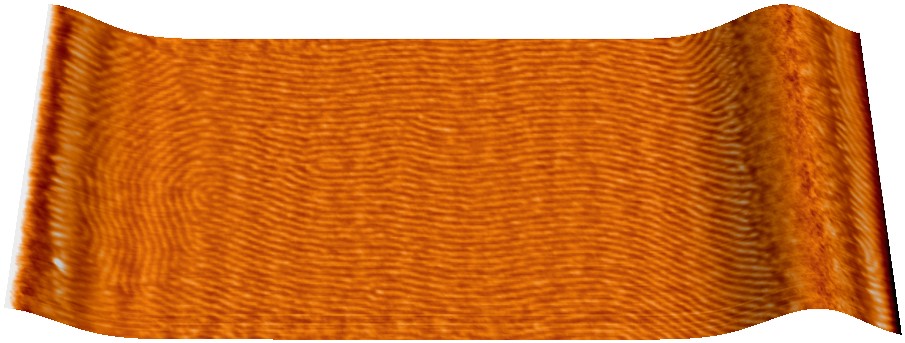}
  \caption{\label{fig:figS3} (a) 3D AFM phase-height image of the BC thin 
film lying on a curved substrate after 3.5 h of thermal annealing at $T$=373K. 
(image size: 2.5 $\mu$m $\times$ 1.0 $\mu$m).}

\end{figure}

Fig. \ref{fig:figS3} emphasizes the coupling between the pattern orientation
and the mean curvature of the substrate. As here the Gaussian curvature is
zero, the substrate is topologically equivalent to flat space. Thus, while for
a strictly 2D system no coupling can be expected, the finite thickness of the
film leads to an interaction that penalizes those configurations that involve
an inter-cylinder elastic distortion.

\subsection{Theory: Symmetry considerations.}

In its own Eigensystem, the shape tensor ${\bf S}$ can be written
as ${\bf S} = \sum_{i=1}^2 k_i {\bf u}_i \otimes {\bf u}_i$, where $k_i$ are
the principal curvatures and ${\bf u}_i$ the corresponding Eigenvectors,
where we can choose $|k_1|>|k_2|$ without loss of generality. If a membrane
or thin film has in-plane order characterized by a director field ${\bf n}$,
the curvature free energy per area no longer has to be rotationally 
symmetric, and it may contain additional terms of the form
$({\bf n} \cdot {\bf S} \cdot {\bf n})$, 
$({\bf n} \cdot {\bf S} \cdot {\bf n})^2$, and
$({\bf n} \cdot {\bf S})^2$.
We write the contribution of these terms to the curvature free energy
per area in the general form
\begin{equation}
F_{\bf n} = A {\bf n} \cdot {\bf S} \cdot {\bf n} 
            - B ({\bf n} \cdot {\bf S} \cdot {\bf n})^2
            - C ({\bf n} \cdot {\bf S})^2.
\end{equation}
The product $({\bf n} \cdot {\bf u}_1)$ defines the angle $\theta$ between 
the director and the direction of largest curvature {\em via}
$({\bf n} \cdot {\bf u}_1)^2 = \cos^2 \theta$. Inserting this and using
${\bf n} = ({\bf n} \cdot {\bf u}_1) {\bf u}_1 + ({\bf n} \cdot {\bf u}_2) {\bf u}_2$
and $({\bf n} \cdot {\bf u}_2)^2 = 1 - ({\bf n} \cdot {\bf u}_1)^2$, we obtain
\begin{eqnarray}
F_{\bf n} &=&   A H - (\frac{3}{2} B+2 C) H^2 + (\frac{1}{2} B+C) K
\\ \nonumber
&& + \Big( \frac{A}{2}  (k_1 - k_2)  
    - (B+C)  (k_1 - k_2) \: H \Big)\: \cos(2 \theta)
\\ \nonumber
&&  - \frac{B}{2}  (H^2 -K) \cos(4 \theta).
\end{eqnarray}
The first term can be absorbed in the spontaneous curvature $c_0$, and the
second two terms in the bending and Gaussian modulus $\kappa_b$ and $\kappa_g$,
respectively. The last two terms give the expression for the anisotropic
curvature free energy per area $F_{\mbox{\tiny ani}}$ in the main text (Eq.\
(1)), with $\kappa'= (B+C)$ and $\kappa''= B$, and $c_0'=A/(B+C)$.

\subsection{Theory: SCFT calculations.}

\subsubsection{Basic equations}

We consider a melt of asymmetric $AB$ diblock copolymer molecules confined in a
volume $V$ between two coaxial cylindrical surfaces of radius $R_1$ and
$R_2=R_1+\epsilon$, where $\epsilon$ is the thickness of the confined film. The
two surfaces preferentially attract $A$-monomers. Dirichlet
boundary conditions are applied in the radial direction and periodic boundary
conditions are applied in the in-plane directions.  Each diblock copolymer
molecule consists of $N$ segments of which  a fraction $f$ forms the majority
block $A$. We assume that $A$ and $B$ segments have the same
statistical segment length $b$. The microscopic concentration operators of
$A$ and $B$ segments at a given point $\mathbf{r}(r,\varphi,z)$ are defined as
\begin{eqnarray}
 \hat{\phi}_A(\mathbf{r}) = \frac{1}{\rho_c} \sum_{j=1}^{n} \int_0^f ds 
   \: \delta (\mathbf{r}-\mathbf{r}_{j}(s)) \\
 \hat{\phi}_B(\mathbf{r}) = \frac{1}{\rho_c} \sum_{j=1}^{n} \int_f^1 ds 
   \: \delta (\mathbf{r}-\mathbf{r}_{j}(s))
\end{eqnarray}
respectively. These concentrations are made dimensionless by dividing by the 
average copolymer density $\rho_c$. 
The interaction potential of the melt is
\begin{eqnarray}
 \frac{\mathcal{H_I}}{k_BT} & = & \rho_c \int d\mathbf{r} 
 \left[ \chi N \hat{\phi}_A(\mathbf{r})\hat{\phi}_B(\mathbf{r}) \right. 
   \\ && 
    \left. + \frac{1}{2} \kappa N 
       \left( \hat{\phi}_A(\mathbf{r}) + \hat{\phi}_B(\mathbf{r}) -1 \right)^2 \right] 
      \nonumber \\ \nonumber
   && + \rho_c \int d\mathbf{r} \: H(\mathbf{r})
   \left[ \Lambda_A^{s,a} N \hat{\phi}_A(\mathbf{r}) 
        + \Lambda_B^{s,a} N \hat{\phi}_B(\mathbf{r})  \right]
\end{eqnarray}
where the Flory-Huggins parameter $\chi$ specifies the repulsion of $A$ and $B$
segments. The second term describes a finite compressibility of
the melt \cite{Helfand1975}, which is fixed to $\kappa N = 25$, similar to
previous work on similar systems \cite{Pike2009, Detcheverry2010}.
The terms $\Lambda_{A,B}^{a,s} \: H(\mathbf{r})$ are
surface fields.  We choose a form

\begin{equation}
 H(\mathbf{r}) = \left\{ 
  \begin{array}{l l}
    (1 + \cos(\pi (r-R_1)/\delta)) & \quad \text{ $R_1 \leq r \leq R_1 + \delta$}\\
    0 & \quad \text{ $R_1 + \delta < r < R_2-\delta$} \\
    (1+\cos(\pi(R_2-r)/\delta)) & \quad \text{ $R_2-\delta \leq r \leq R_2$}
  \end{array} \right.
\end{equation}
with $\delta = 0.2 R_g$. The value of $\delta$ must be chosen
small enough relative to the domain size so that its finite size does not
affect the phase behavior of the thin films significantly.
$\Lambda_{A,B}^{s,a}$ gives the strength of the interaction between block $A$
or $B$, respectively, and the substrate ($s$) and air ($a$) interface. The
``surface interaction energies per area'' of component $A$ or $B$ are defined
as the integrated surface energy per area of a hypothetical film of $A$ or $B$
monomers with density $\hat{\phi}_{A,B} \equiv 1$, i.e., $\gamma_{A,B}^{s,a} =
\rho_c \int d\mathbf{r} \: H(\mathbf{r}) \Lambda_{A,B}^{s,a}$.  They will be
given in units of $\hat{\gamma} = \rho_c R_g  k_BT$ (which is a unit of energy
per area).  In the following, we shall set $k_BT=1$, for notational simplicity. 

In the membrane study we assume that the two surfaces are symmetric
for each block, $\Lambda_A^a N = -120$ and $\Lambda_B^a N = -115$
corresponding to $\gamma_A N = -24 \hat{\gamma}$ and $\gamma_B N = -23 \hat{\gamma}$. 
In the curved supported thin films, we choose
symmetric surface interactions for the $B$-block $\Lambda_B^s N = \Lambda_B^a N =
-30$ corresponding to $\gamma_B N = -6 \hat{\gamma}$, 
and asymmetric conditions for the $A$-block. Specifically, we study
two cases: 
\begin{itemize}
\item[I.] The substrate attracts the $A$-block more strongly than the free
(air) surface ($\Lambda_A^s N = -120, \Lambda_A^a N = -50$) with corresponding
surface energy per area $\gamma_A^s N = -24 \hat{\gamma}$, 
$\gamma_A^a N= - 10 \hat{\gamma}$.  
\item[II.] The free surface attracts the $A$-block more strongly
than the substrate ($\Lambda_A^s = -50, \Lambda_A^a N = -100$) with
corresponding surface energy per area $\gamma_A^s N= -10 \hat{\gamma}$ and
$\gamma_A^a N= -20 \hat{\gamma}$. 
\end{itemize}
Our calculations are done in the grand canonical ensemble with the free energy
\begin{eqnarray}
 F_{GC} & = & 
   -e^{\mu}Q + \rho_c \int d\mathbf{r} 
  \left[\chi N\phi_A(\mathbf{r})\phi_B(\mathbf{r}) \right.
\nonumber \\ &&
    \quad  + \left. \frac{\kappa}{2}\left(\phi_A(\mathbf{r})
   +\phi_B(\mathbf{r})-1 \right)^2\right] \nonumber \\
 &  & - \rho_c \int d\mathbf{r} 
  \left[ \omega_A(\mathbf{r})\phi_A(\mathbf{r})
   +\omega_B(\mathbf{r})\phi_B(\mathbf{r}) \right] \nonumber \\
 &  &  + \rho_c \int d\mathbf{r} H(\mathbf{r})  N
   \left[ \Lambda_A^{s,a} \phi_A(\mathbf{r}) + \Lambda_B^{s,a} \phi_B(\mathbf{r}) \right]
\label{eq:fgc}
\end{eqnarray}
where $\mu$ is chemical potential, $Q$ the partition function of a single 
non-interacting polymer chain, 
\begin{equation}
 Q = \int d\mathbf{r} q(\mathbf{r},s) q^{\dag}(\mathbf{r},1-s)
\end{equation}
and $q(\mathbf{r},s)$ and $q^{\dag}(\mathbf{r},1-s)$ satisfy the modified 
diffusion equation
\begin{eqnarray}
 \frac{\partial q(\mathbf{r},s)}{\partial s} 
  = \Delta q(\mathbf{r},s) -\omega_{\alpha}(\mathbf{r},s) q(\mathbf{r},s) 
\end{eqnarray}
with 
\begin{equation}
 \omega_{\alpha}(\mathbf{r},s) = \left\{ 
  \begin{array}{l l}
    \omega_A(\mathbf{r}) & \quad \text{for $0<s<f$}\\
    \omega_B(\mathbf{r}) & \quad \text{for $f<s<1$}
  \end{array} \right.
\end{equation}
and the initial condition $q(\mathbf{r},0) = 1$. The diffusion equation for
$q^{\dag}(\mathbf{r},1-s)$ is similar with $\omega_\alpha (\mathbf{r},s)$
replaced by $\omega_\alpha (\mathbf{r},1-s)$ and the same initial condition,
$q^{\dag}(\mathbf{r},0)=1$.
By finding the extremum of the free energy, Eq.\ (\ref{eq:fgc}) with respect to
$\omega_{A,B}(\mathbf{r})$ and $\phi_{A,B}(\mathbf{r})$, we get the
self-consistent equations,
\begin{eqnarray}
\frac{\omega_A(\mathbf{r})}{N} & = & \chi \phi_B(\mathbf{r}) 
   + \kappa \left[\phi_A(\mathbf{r})+\phi_B(\mathbf{r})-1\right] 
   + \Lambda_A H(\mathbf{r}) \nonumber \\
\frac{\omega_B(\mathbf{r})}{N} & = & \chi \phi_A(\mathbf{r}) 
   + \kappa \left[\phi_A(\mathbf{r})+\phi_B(\mathbf{r})-1 \right] 
   + \Lambda_B H(\mathbf{r}) \nonumber \\
 \phi_A(\mathbf{r}) & = & 
   \frac{1}{\rho_c} \: {\rm e}^\mu
   \int_0^f ds \: q(\mathbf{r},s) \: q^{\dag}(\mathbf{r},1-s) \nonumber \\
 \phi_B(\mathbf{r}) & = & 
   \frac{1}{\rho_c} \: {\rm e}^\mu
   \int_f^1 ds \: q(\mathbf{r},s)\: q^{\dag}(\mathbf{r},1-s)
\end{eqnarray} 

\subsubsection{Boundary conditions}

The SCFT calculations are done in cylindrical coordinates $(r,\phi,z)$, where $r$
is the direction normal to the film or membrane surface and $z$ is the
direction of zero curvature.  Configurations $C_{\parallel}$, $C_\perp$ with
cylinder orientations parallel or perpendicular to the direction of curvature
can be obtained with periodic boundary conditions in the ($\phi,z$) directions.
In order to impose a given tilted orientation with tilt angle $\theta$ (as in
Fig.\ \ref{fig:fig3n}c, inset), we must apply tilted boundary conditions,
either in the $z$ or in the $\phi$ direction. We do this by using affine
coordinates $(r,u,v)$ with periodic boundary conditions in $(u,v)$ and (1) $u =
\varphi, v = z-\varphi a$, $a=R \tan \theta$ or (2) $(r,u,v)$, $u=\varphi - b
z, v=z$, $b = 1/(R\tan \theta$). In case (1), $a=0$ corresponds to the
perpendicular configuration $C_\perp$ ($\theta = 0$), and in case (2) $b=0$
corresponds to the parallel configuration $C_\parallel$ ($\theta = \pi/2$).  

We solve the modified diffusion equations with periodic boundary conditions in
effectively two dimensions: $(r,v)$, independent of $u$ in case (1), and
$(r,u)$, independent of $v$ in case (2). This enforces tilted orientations of
cylinders.  In general, the Laplace-Beltrami operator has the following form 
\begin{eqnarray}
 \Delta_{LB} = \frac{1}{\sqrt{|\det g|}} 
   \sum_{ij} \frac{\partial}{\partial x_i} \left( g^{ij} \sqrt{|\det g|} \frac{\partial}{\partial x_j}\right)
\end{eqnarray}
where $g_{ij}$ is the metric tensor and $g^{ij}$ its inverse. The Laplace-Beltrami operator in cases (1) and (2) is thus given by
\begin{eqnarray}
 \Delta_{LB}^{(1)} 
     & = & \frac{1}{r} \frac{\partial}{\partial r} + \frac{\partial^2}{\partial r^2} 
             + \frac{1}{r^2} \: \frac{\partial^2}{\partial u^2} 
             - \frac{2a}{r^2} \frac{\partial^2}{\partial u \: \partial v} \nonumber \\
             &  & + \left( 1 + \frac{a^2}{r^2} \right) \frac{\partial^2}{\partial v^2} 
                  \qquad \mbox{case (1)}\\
 \Delta_{LB}^{(2)} & = & \frac{1}{r} \frac{\partial}{\partial r} + \frac{\partial^2}{\partial r^2}
                         + \left( \frac{1}{r^2} + b^2 \right) \frac{\partial^2}{\partial u^2} \nonumber \\
                   &  & -2b \frac{\partial^2}{\partial u \: \partial v} + \frac{\partial^2}{\partial v^2}
\qquad \mbox{case (2)}.
\end{eqnarray}

The modified diffusion equations were solved using the the
Crank-Nicolson method. We used the setup (1) for small angles $0 < \theta <
\pi/4$ and the setup (2) for large angles $ \pi/4 < \theta < \pi/2$, and
compared the results from both setups at the angle $\theta = \pi/4$, to verify
that both setups give the same result.

\subsubsection{Discretization errors and correction}

\begin{figure}[h]
\centering
\includegraphics[width=7.5cm]{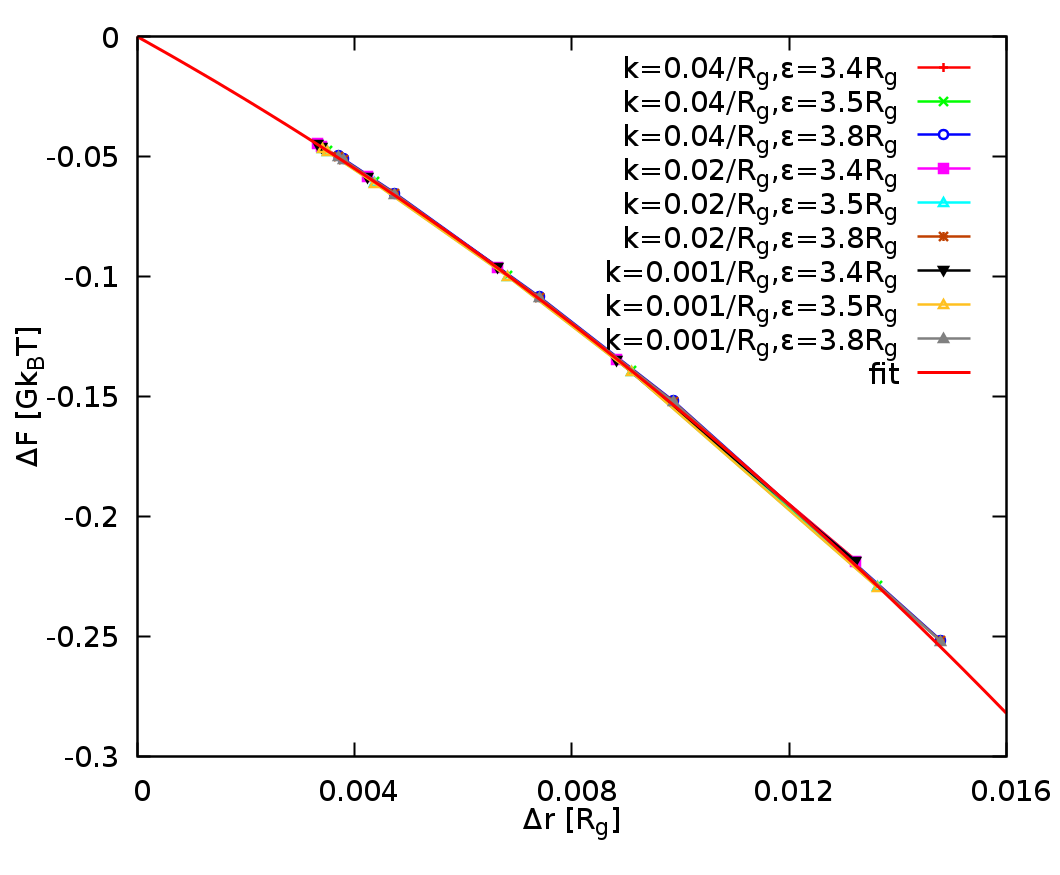}
  \caption{\label{fig:figS4} Shift of free energy $\Delta F$
as a function of discretization $\Delta r$ for symmetric films with surface
interactions $\Lambda_A^a N = -120, \Lambda_B^a N = -115$, for different curvatures
and film thicknesses $\epsilon$ as indicated. Solid line: fit function 
$f(x)= -12.6x-277x^2-2367x^3$.}
\end{figure}

\begin{figure}[h]
\centering
\includegraphics[width=7.5cm]{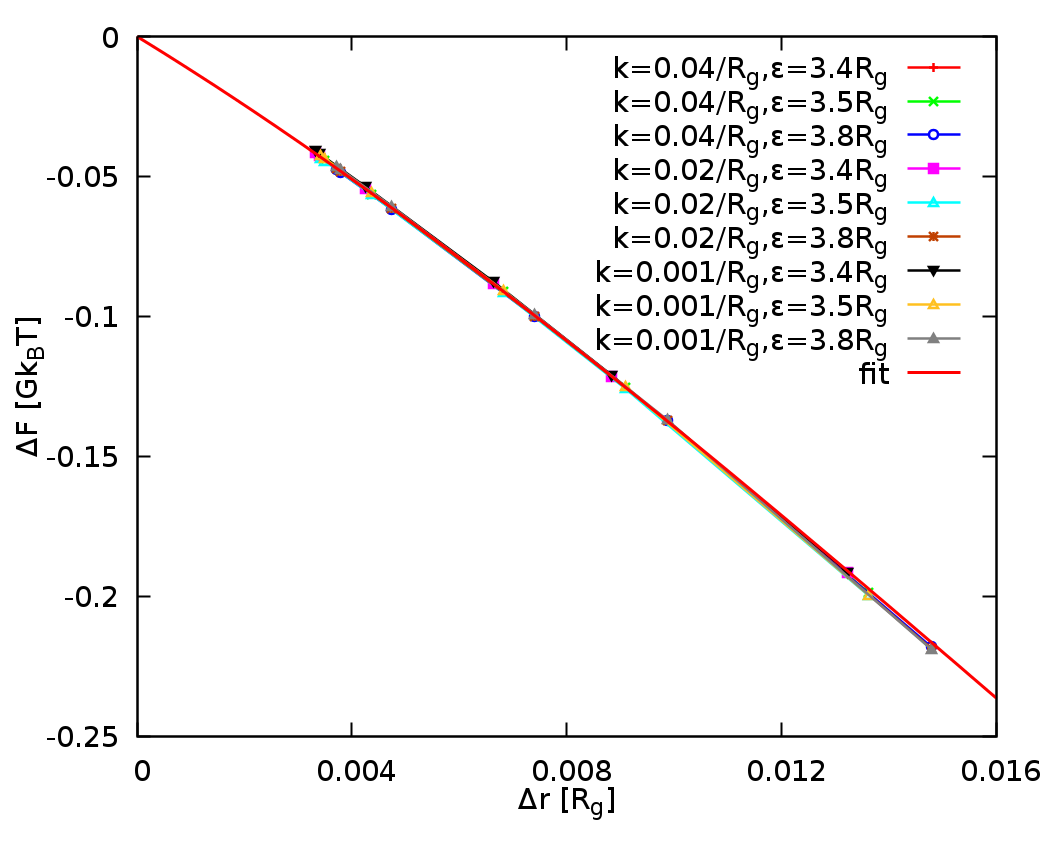}
  \caption{\label{fig:figS5} Shift of free energy $\Delta F$ as a function of discretization
$\Delta r$ for films with asymmetric surface interactions $\Lambda_A^s N =
-120, \Lambda_B^s N = -30$ at the fixed substrate and $\Lambda_A^a N = -50,
\Lambda_B^a N = -30$ at the free surface for different curvatures and film 
thicknesses $\epsilon$  as indicated. Solid line: fit function 
$f(x)=-11.75x-270x^2+5035x^3$.}
\end{figure}

\begin{figure}[h]
\centering
\includegraphics[width=7.5cm]{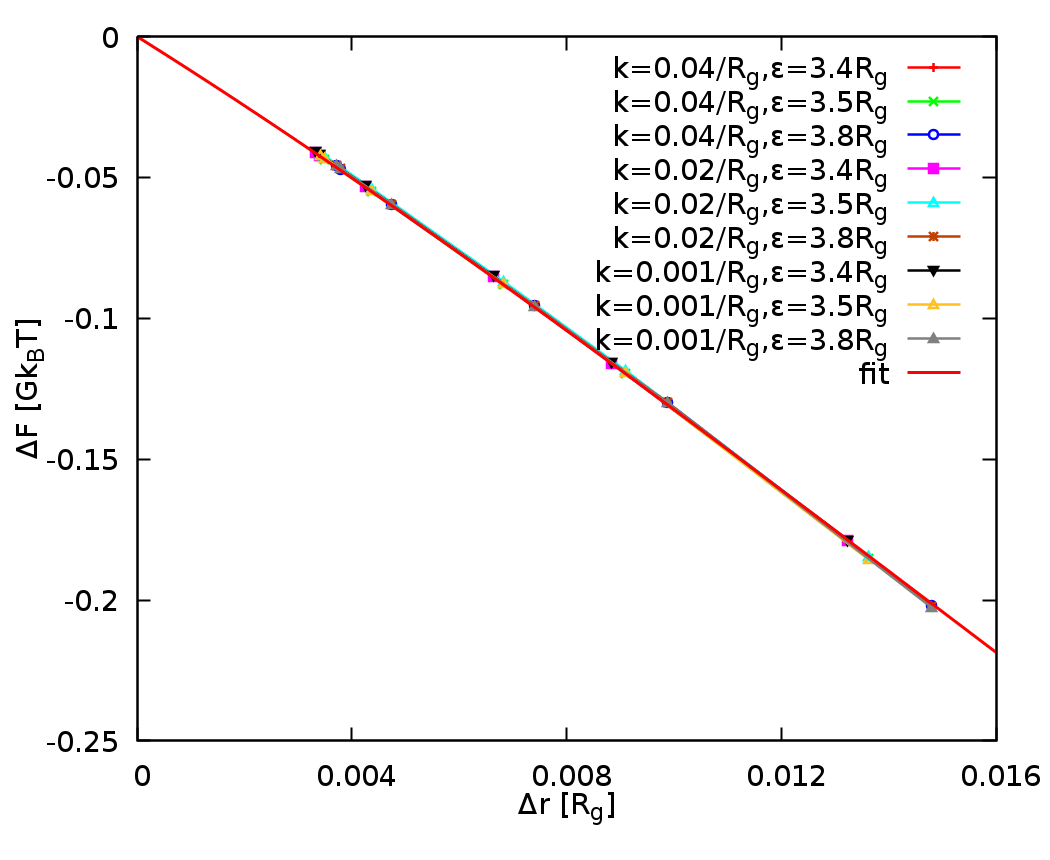}
  \caption{\label{fig:figS6} 
Shift of free energy $\Delta F$ as a function of discretization
$\Delta r$ for films with asymmetric surface interactions $\Lambda_A^s N =
-50, \Lambda_B^s N = -30$ at the fixed substrate and $\Lambda_A^a N = -100,
\Lambda_B^a N = -30$ at the free surface for different curvatures and film 
thicknesses $\epsilon$  as indicated. Solid line: fit function 
$f(x)=-11.95x-167x^2+3737x^3$.}
\end{figure}

The discretizations in the azimuthal and thin film directions ($z$ and $r$,
respectively) were chosen as $\Delta z=0.05R_g$ and $\Delta r =0.01R_g$, and the
parameter $s$ was discretized in steps of $\Delta s = 0.0001$. Whereas most of
the choices are not critical, we found that the discretization in the $r$ direction
has a significant influence on the resulting free energies, and discretization
errors could not be neglected. On the other hand, we also found that they lead
to an energy shift $\Delta F$ which depends only on $\Delta r$ and not on the
film thickness or curvature. We therefore studied the dependence of
$\Delta F$ on $\Delta r$ systematically for different values of the film
thickness and curvature. Then we fitted the result to a third order polynomial,
resulting in the estimate $\Delta F (\Delta r) = -11.75 \Delta r - 270 \Delta
r^2 + 5035 \Delta r^3 $ ($\Delta F (\Delta r) = -11.95 \Delta r - 167 \Delta
r^2 + 3737 \Delta r^3 $) and $\Delta F (\Delta r) = -12.6 \Delta r - 277 \Delta
r^2 - 2367 \Delta r^3 $ with asymmetric and homogeneous surface interactions,
respectively. Figs.\ \ref{fig:figS4}, \ref{fig:figS5} and \ref{fig:figS6}
present the fitting results. These corrections were then applied to the 
results of the SCFT calculations.


\subsubsection{Optimum thickness of free-standing membranes}

\begin{figure}[h]
\centering
\includegraphics[width=8cm]{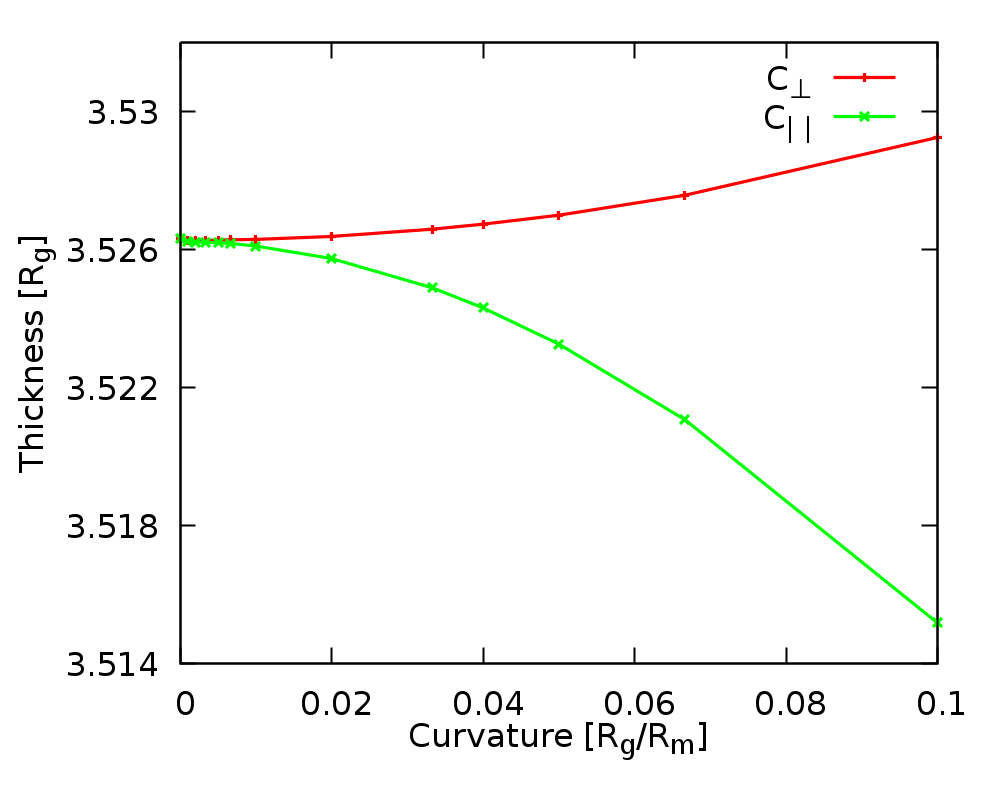}
  \caption{\label{fig:figS7} The optimal film thickness of
free-standing membranes as a function of curvature. }
\end{figure}

Fig. \ref{fig:figS7} shows the behavior of the optimum film thickness of
free-standing membranes as a function of curvature both parallel and
perpendicular configurations.  Interestingly, the optimal thickness is not
affected strongly by curvature in the range of curvatures considered here.

\clearpage

\end{document}